# Enhancement of nucleate boiling by combining the effects of surface structure and mixed wettability: A lattice Boltzmann study

W. X. Li, Q. Li[*], Y. Yu, and Z. X. Wen

*School of Energy Science and Engineering, Central South University, Changsha 410083, China*

*Corresponding author: qingli@csu.edu.cn

**Abstract**

The combination of microstructures and mixed wettability for enhancing nucleate boiling has attracted much attention in recent years. However, in the existing experimental and numerical studies, the tops of microstructures are entirely subjected to wettability modification, which makes the influences of mixed wettability dependant on the characteristic length of microstructures. In order to disclose the joint effects of surface structure and mixed wettability on nucleate boiling, in this work we propose an improved type of pillar-textured surface with mixed wettability, in which the tops of square pillars are *partially* subjected to wettability modification. Numerical investigation of the boiling heat transfer performance on the improved mixed-wettability surface is carried out using a three-dimensional thermal multiphase lattice Boltzmann model. The numerical results show that the width of the wettability-modified region plays an important role in the boiling performance of the improved mixed-wettability surface and the best boiling performance is achieved in the situation that the width of the wettability-modified region is sufficiently large but the bubble nucleated on the pillar top still does not interfere with the coalescence-departure mechanism of the bubbles nucleated around the pillar, which optimizes the joint effects of surface structure and mixed wettability for enhancing nucleate boiling. The influences of the shape of the wettability-modified region are also studied. Among the investigated shapes, the square is found to perform better than the other two shapes.

**Keywords:** Nucleate boiling; mixed wettability; boiling enhancement; lattice Boltzmann

**Nomenclature**

| | |
|---|---|
| $a, b$ | parameters in the non-ideal equation of state |
| $c$ | lattice speed |
| $c_s$ | lattice sound speed |
| $c_V$ | specific heat at constant volume |
| $\mathbf{e}$ | discrete particle velocity |
| $f$ | density distribution function |
| $\mathbf{F}$ | total force exerted on the system |
| $\mathbf{F}_b$ | buoyant force |
| $\mathbf{F}_m$ | intermolecular interaction force |
| $g$ | gravitational acceleration |
| $H$ | pillar height |
| $h$ | heat flux |
| $\bar{h}$ | normalized heat flux |
| $L_x$ | length of the computational domain |
| $L_y$ | width of the computational domain |
| $L_z$ | height of the computational domain |
| l.u. | lattice units |
| $p_c$ | critical pressure |
| $p_{\mathrm{EOS}}$ | non-ideal equation of state |
| $R$ | gas constant |
| $S$ | area of the wettability-modified region |
| $t$ | time |
| $T_c$ | critical temperature |
| $T_s$ | saturation temperature |
| $\mathbf{u}$ | macroscopic velocity |
| $W$ | pillar width |
| $W_{\mathrm{mod}}$ | width of the wettability-modified region |
| $w$ | weighting coefficients |

*Greek symbols*

$\omega$      acentric factor

$\lambda$      thermal conductivity

$\rho$      macroscopic density

$\theta$      contact angle

$\nu$      kinematic viscosity

$\delta_t$      time step

$\chi$      thermal diffusivity

$\psi$      pseudopotential

$\Delta T$      wall superheat

*Subscripts or superscripts*

$\alpha$      lattice velocity direction

$c$      critical

$eq$      equilibrium

$L$      liquid

mod      wettability-modified region

phi      hydrophilic

pho      hydrophobic

$V$      vapor

# 1. Introduction

Nucleate boiling is recognized as a highly efficient mode of heat transfer [1] and has been widely utilized in various energy conversion and heat exchange systems, such as power generation, electronic cooling, propulsion, chemical processes, etc. [1-3]. In the past decades, many researchers have carried out either experimental or numerical studies to investigate the mechanism of nucleate boiling and explore how to improve the boiling heat transfer performance. Nevertheless, owing to the extreme complexity of the nucleate boiling process, the mechanism of enhancing nucleate boiling heat transfer has not been well understood. In recent years, two methods of surface modification have attracted much attention in the studies of enhancing nucleate boiling heat transfer: one is to regulate and control the wettability of the heating surface [4, 5] and the other is applying micro/nano-scale structures [6].

In some early studies of boiling heat transfer, the importance of surface wettability was not recognized because the wettability effect was included in the effects of surface characteristics such as roughness [7]. Through numerous experiments, it is now widely recognized that the surface wettability has significant influences on the boiling curve and the critical heat flux of boiling heat transfer. Generally, a hydrophobic surface can more easily trigger the nucleation of bubbles than a hydrophilic surface because decreasing the surface wettability can reduce the energy barrier required for the liquid-vapor phase change [8, 9]. Accordingly, the onset of boiling on a hydrophobic surface begins at a lower wall superheat. However, the critical heat flux decreases dramatically on a hydrophobic surface, while a hydrophilic surface shows a higher value of the critical heat flux [10]. Therefore, a recent trend in surface modification is to make full use of the advantages of hydrophobic and hydrophilic surfaces, which leads to the applications of heterogeneous surfaces combining hydrophilicity and hydrophobicity for enhancing nucleate boiling through both experiments and numerical studies.

Betz *et al*. [11] manufactured oxidized silicon wafers surfaces with networks combining hydrophilic and hydrophobic regions and experimentally demonstrated that the heterogeneous surfaces perform better than a hydrophilic surface with 7° wetting angle. Jo *et al*. [12] coated hydrophobic Teflon dots on a hydrophilic silicon surface in the absence of micro-scale roughness and showed that the heterogeneous surface can provide better nucleate boiling heat transfer than the surfaces with homogeneous wettability. They found that the number of hydrophobic dots and the pitch distance between dots have important influences on the boiling performance. Subsequently, Jo *et al*. [13] fabricated a heterogeneous wetting surface comprising a wetting pattern located at the head of surface microstructures. They observed that the heterogeneous wetting surface with micro-pillars can interrupt the expansion and coalescence of bubbles and hence enhance the boiling heat transfer performance.

Recently, Zhang *et al*. [14] and Shen *et al*. [15] have also experimentally studied boiling heat transfer by applying the combination of microstructures and mixed wettability. In the work of Zhang *et al*. [14], they observed that the combination of microstructures with hydrophobicity contributes to a high heat transfer coefficient at low heat fluxes and leads to a lower wall superheat for boiling onset as the microstructures provide potential bubble nucleation sites and the hydrophobicity can reduce the energy barrier of the liquid-vapor phase change. Shen *et al*. [15] fabricated hybrid wetting surfaces with square pillars by the chemical deposition approach. They found that the hybrid wetting surfaces are superior to spatially homogeneous wetting surfaces and can provide higher values of the critical heat flux under the same geometric size and heating power conditions. Among the investigated hybrid modes, they observed that the hybrid mode of which the tops of square pillars are hydrophobic whereas the bottom substrate is hydrophilic is the most effective one.

Besides the aforementioned experimental studies, some numerical studies of boiling heat transfer on heterogeneous wetting surfaces have been performed using the lattice Boltzmann (LB) method, which is a mesoscopic method built on the kinetic Boltzmann equation [16] and has been developed into an efficient numerical approach for simulating fluid flow [17-21] and heat transfer [22-24]. In recent years, the LB method has been applied to simulate boiling and evaporation [25-31]. Most of these studies are based on the pseudopotential multiphase LB method [32, 33] together with a thermal LB model or a solver for simulating the temperature field. A distinctive advantage of the pseudopotential multiphase LB method for simulating interface phenomena is that the interface between different phases can arise, deform, and migrate naturally without using any techniques to track or capture the interface [22], which is usually required in conventional numerical methods. Moreover, the implementation of wetting boundaries (contact angles) in the pseudopotential multiphase LB method is much simpler than that in conventional numerical methods, which makes it very suitable for simulating multiphase flows involving complex surfaces, such as microstructured surfaces [34, 35].

Inspired by the experimental study of Betz *et al*. [11], Gong and Cheng [36] have investigated the boiling heat transfer on flat surfaces with mixed wettability using a phase-change model based on the pseudopotential multiphase LB method. The effects of the size of hydrophobic spots and the pitch distance between the spots on the boiling performance were studied. They numerically demonstrated that adding hydrophobic spots on smooth hydrophilic surfaces can promote bubble nucleation and reduce the nucleation time. Moreover, Li *et al*. [37, 38] have numerically investigated the boiling heat transfer performance on a type of hydrophilic-hydrophobic mixed surface, which is textured with pillars consisting of hydrophilic side walls and hydrophobic tops. They found that the hydrophobicity of the tops of pillars can promote bubble nucleation and reduce the required wall superheat for boiling onset. They numerically showed that decreasing the wettability of the tops of pillars causes a leftward shift of the boiling curve and a leftward-upward shift of the heat transfer coefficient curve. Recently, Ma *et al*. [39] investigated the boiling performances on four types of micro-pillar heat sinks with mixed-wettability patterns. They found that the heat sink with hydrophobic pillar tops and hydrophilic base provides the best boiling heat transfer performance. Furthermore, Ma and Cheng [40] have numerically studied the boiling heat transfer on micro-pillar and micro-cavity hydrophilic heaters and showed that the high critical heat flux of the micro-pillar structured hydrophilic heater is mainly attributed to the considerable effect of capillary wicking.

Although significant progress has been achieved in applying the combination of microstructures and mixed wettability for boiling heat transfer, an important limitation of current studies should be pointed out, i.e., in the existing experimental and numerical studies, the tops of microstructures are *entirely* subjected to surface modification, which means that in these studies the effect of mixed wettability is non-independent and related to the characteristic length of the microstructures (e.g., the width of the microstructures). In order to disclose the joint effects of surface structure and mixed wettability, in the present study we numerically investigate the boiling heat transfer performance on an improved type of pillar-textured surface with mixed wettability, in which the tops of square pillars are *partially* subjected to wettability modification. To the best of the authors' knowledge, such an improved mixed-wettability surface has never been studied previously, neither experimentally nor numerically. Numerical simulations in the present work are performed using a three-dimensional (3D) thermal multiphase LB model with liquid-vapor phase change [38]. The rest of this paper is organized as follows. The 3D thermal multiphase LB model with liquid-vapor phase change is introduced in Section 2. In Section 3, numerical investigations are presented and some discussions of the joint effects of surface structure and mixed wettability for enhancing nucleate boiling are also provided there. Finally, a brief summary is given in Section 4.

## 2. Numerical model

In this section, we briefly introduce the 3D thermal multiphase LB model with liquid-vapor phase change. In the LB method, the fluid flow is simulated by solving the discrete Boltzmann equation with certain collision operator, such as the Bhatnagar-Gross-Krook (BGK) [41, 42] collision operator and the multiple-relaxation-time (MRT) [43, 44] collision operator. The BGK collision operator can be viewed as a special case of the MRT collision operator. Using the MRT collision operator, the LB equation can be written as follows [22, 38, 45]:

$$f_\alpha\left(\mathbf{x}+\mathbf{e}_\alpha \delta_t, t+\delta_t\right)=f_\alpha(\mathbf{x}, t)-\left.\bar{\Lambda}_{\alpha\beta}\left(f_\beta-f_\beta^{eq}\right)\right|_{(\mathbf{x},t)}+\left.\delta_t\left(G_\alpha-0.5\bar{\Lambda}_{\alpha\beta}G_\beta\right)\right|_{(\mathbf{x},t)}, \qquad (1)$$

where $f_\alpha$ is the density distribution function, $f_\alpha^{eq}$ is the equilibrium distribution, $\mathbf{x}$ is the spatial position, $\mathbf{e}_\alpha$ is the discrete velocity, $t$ is the time, $\delta_t$ is the time step, $\bar{\Lambda}_{\alpha\beta}=\left(\mathbf{M}^{-1}\Lambda\mathbf{M}\right)_{\alpha\beta}$ is the collision operator, in which $\mathbf{M}$ is a transformation matrix and $\Lambda$ is a diagonal matrix, and $G_\alpha$ is the forcing term in the discrete velocity space.

The three-dimensional nineteen-velocity (D3Q19) lattice model is employed, which corresponds to the following lattice velocities [38]:

$$\mathbf{e}_\alpha = \begin{bmatrix} 0 & 1 & -1 & 0 & 0 & 0 & 0 & 1 & -1 & 1 & -1 & 1 & -1 & 1 & -1 & 0 & 0 & 0 & 0 \\ 0 & 0 & 0 & 1 & -1 & 0 & 0 & 1 & -1 & -1 & 1 & 0 & 0 & 0 & 0 & 1 & -1 & 1 & -1 \\ 0 & 0 & 0 & 0 & 0 & 1 & -1 & 0 & 0 & 0 & 0 & 1 & -1 & -1 & 1 & 1 & -1 & -1 & 1 \end{bmatrix}. \quad (2)$$

Using the transformation matrix $\mathbf{M}$, the right-hand side of Eq. (1) can be implemented as follows:

$$\mathbf{m}^* = \mathbf{m} - \mathbf{\Lambda}(\mathbf{m} - \mathbf{m}^{eq}) + \delta_t \left(\mathbf{I} - \frac{\mathbf{\Lambda}}{2}\right)\mathbf{S}, \quad (3)$$

where $\mathbf{m} = \mathbf{Mf}$, $\mathbf{m}^{eq} = \mathbf{Mf}^{eq}$, $\mathbf{I}$ is the unit matrix, and $\mathbf{S} = \mathbf{MG}$ is the forcing term in the moment space. The details of the transformation matrix $\mathbf{M}$, the equilibria $\mathbf{m}^{eq}$, the diagonal matrix $\mathbf{\Lambda}$ for relaxation times, and the forcing term $\mathbf{S}$ can be found in Ref. [38]. The streaming process is given by

$$f_\alpha(\mathbf{x} + \mathbf{e}_\alpha \delta_t, t + \delta_t) = f_\alpha^*(\mathbf{x}, t), \quad (4)$$

where $\mathbf{f}^* = \mathbf{M}^{-1}\mathbf{m}^*$ and $\mathbf{M}^{-1}$ is the inverse matrix of the transformation matrix. The macroscopic density and velocity can be obtained by the following relationships:

$$\rho = \sum_\alpha f_\alpha, \quad \rho \mathbf{u} = \sum_\alpha \mathbf{e}_\alpha f_\alpha + \frac{\delta_t}{2}\mathbf{F}, \quad (5)$$

where $\mathbf{F}$ is the total force exerted on the system.

For single-component multiphase flows, the intermolecular interaction force in the pseudopotential multiphase LB method is given by [22, 38, 46, 47]

$$\mathbf{F}_m = -G\psi(\mathbf{x})\sum_\alpha w_\alpha \psi(\mathbf{x} + \mathbf{e}_\alpha \delta_t)\mathbf{e}_\alpha, \quad (6)$$

where $\psi(\mathbf{x})$ is the pseudopotential, $G$ is the interaction strength, and $w_\alpha$ are the weights. For the D3Q19 lattice, the weights $w_\alpha$ in Eq. (6) are given by $w_{1-6} = 1/6$ and $w_{7-18} = 1/12$. The total force of the system includes the interaction force $\mathbf{F}_m$ and the buoyant force given by $\mathbf{F}_b = (\rho - \rho_{ave})\mathbf{g}$, where $\rho_{ave}$ is the average density in the computational domain and $\mathbf{g} = (0, 0, -g)$ is the gravitational acceleration.

The pseudopotential $\psi(\mathbf{x})$ is taken as $\psi(\mathbf{x}) = \sqrt{2(p_{EOS} - \rho c_s^2)/Gc^2}$ [48, 49], in which $p_{EOS}$ is the non-ideal equation of state and $c=1$ is the lattice speed. In the present study, the Peng-Robinson equation of state is utilized [49], i.e.,

$$p_{\text{EOS}} = \frac{\rho RT}{1-b\rho} - \frac{a\varphi(T)\rho^2}{1+2b\rho-b^2\rho^2}, \tag{7}$$

where $\varphi(T) = \left[1+\left(0.37464+1.54226\omega-0.26992\omega^2\right)\left(1-\sqrt{T/T_c}\right)\right]^2$, $a = 0.45724R^2T_c^2/p_c$, and $b = 0.0778RT_c/p_c$. The parameter $\omega = 0.344$ is the acentric factor. When the viscous heat dissipation is neglected, the governing equation for the temperature field is given by [27, 38, 50]

$$\partial_t T = -\mathbf{u}\cdot\nabla T + \frac{1}{\rho c_V}\nabla\cdot(\lambda\nabla T) - \frac{T}{\rho c_V}\left(\frac{\partial p_{\text{EOS}}}{\partial T}\right)_\rho \nabla\cdot\mathbf{u}, \tag{8}$$

where $\lambda$ is the thermal conductivity and $c_V$ is the specific heat at constant volume. Similar to the previous studies in Refs. [27, 37, 38], the present study also solves the temperature equation (8) with the classical fourth-order Runge-Kutta scheme for the time discretization. The isotropic central schemes are applied for the spatial discretization. For a quantity $\phi$, the spatial gradient of $\phi$ and the Laplacian of $\phi$ can be calculated by the following second-order isotropic difference schemes [18], respectively

$$\partial_i\phi(\mathbf{x}) \approx \frac{1}{c_s^2\delta_t}\sum_\alpha \omega_\alpha \phi(\mathbf{x}+\mathbf{e}_\alpha\delta_t)e_{\alpha i}, \tag{9}$$

$$\nabla^2\phi(\mathbf{x}) \approx \frac{2}{c_s^2\delta_t^2}\sum_\alpha \omega_\alpha\left[\phi(\mathbf{x}+\mathbf{e}_\alpha\delta_t)-\phi(\mathbf{x})\right], \tag{10}$$

where $c_s = c/\sqrt{3}$ is the lattice sound speed. The weights $\omega_\alpha$ are given by $\omega_{1-6} = 1/18$ and $\omega_{7-18} = 1/36$. The saturation temperature is taken as $T_s = 0.8T_c$ (here $T_c$ is the critical temperature), which corresponds to the liquid density $\rho_L \approx 7.2$ and the vapor density $\rho_V \approx 0.197$. In the present work all the quantities are taken in the lattice units, i.e., the units in the LB method.

## 3. Numerical results and discussions

### 3.1. The simulation setup

The sketch of textured heating surfaces with micro-pillars is shown in Fig. 1 and the domain within the red dotted lines is chosen as the computation domain owing to employing the periodic boundary condition in the $x$ and $y$ directions. The grid system of the computational domain is taken as $L_x \times L_y \times L_z = 150\,\text{l.u.} \times 150\,\text{l.u.} \times 250\,\text{l.u.}$, in which l.u. represents lattice units. Figure 2 illustrates three types of pillar-textured surfaces with different wettability, i.e., a homogeneous hydrophilic surface, a mixed-wettability surface whose pillars consist of hydrophilic side walls and hydrophobic tops, and an improved mixed-wettability surface in which the tops of pillars are partially subjected to wettability modification (denoted by the light blue region). The square pillar is located at the center of the bottom substrate and the height and width of the square pillar are denoted by $H$ and $W$, respectively.

Initially, the computational domain is a liquid phase ($0 \leq z < 150\,\text{l.u.}$) below its saturated vapor phase ($150\,\text{l.u.} \leq z \leq L_z$) and the initial temperature of the computational domain is taken as $T_s = 0.8 T_c$. The surfaces of the square pillar and the bottom of the computational domain are the heating surfaces and the temperature is given by $T_b = T_s + \Delta T$, while the temperature of the top boundary is maintained at $T_s$. The periodic boundary condition is employed in the $x$ and $y$ directions (the numerical implementation can be found in Sec. 4.4.1 of Ref. [51]) and the Zou-He boundary scheme [52] is applied at the heating surfaces. In our simulations, the thermal conductivity is given by

$$\lambda = \lambda_V \frac{\rho_L - \rho}{\rho_L - \rho_V} + \lambda_L \frac{\rho - \rho_V}{\rho_L - \rho_V}, \tag{11}$$

where $\lambda_V$ is defined as $\lambda_V = \rho_V c_V \chi_V$, in which $\chi_V$ is taken as 0.03 and $c_V = 10$. The ratio $\lambda_L / \lambda_V$ is chosen as 15 and the gravitational acceleration is set to $g = 3 \times 10^{-5}$.

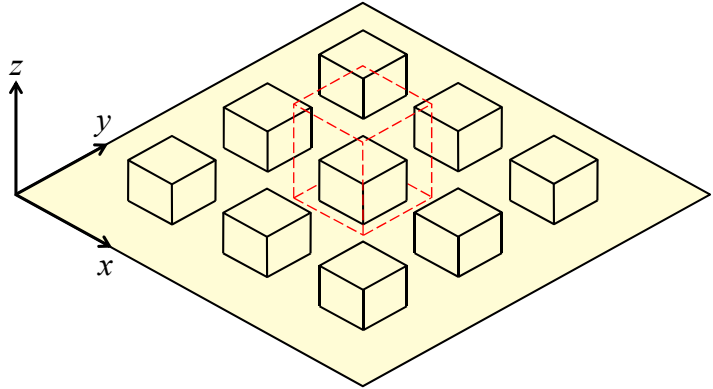

**Fig. 1.** Schematic illustration of pillar-textured surfaces.

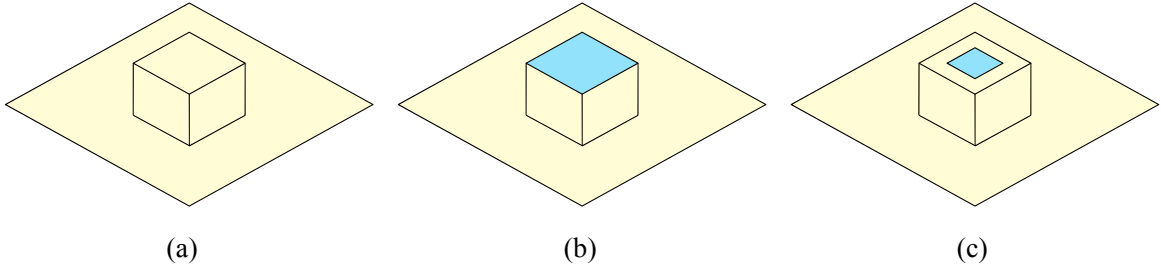

**Fig. 2.** Schematic illustration of three types of pillar-textured surfaces with different wettability. (a) a homogeneous hydrophilic surface, (b) a mixed-wettability surface whose pillars consist of hydrophilic side walls and hydrophobic tops, and (c) an improved mixed-wettability surface in which the tops of pillars are partially subjected to wettability modification (the light blue region).

### 3.2. Boiling performance on the improved mixed-wettability surface

In this section, the boiling performance on the improved mixed-wettability surface is investigated. In simulations, the height of the square pillar is taken as $H = 20$ l.u. and the contact angles of the hydrophilic and hydrophobic regions are chosen as $\theta_{phi} \approx 37^{\circ}$ and $\theta_{pho} \approx 94^{\circ}$, respectively. Numerical implementation of the contact angles can be found in Ref. [53]. For comparison, we briefly display the basic characteristics of boiling on the homogeneous hydrophilic surface and the mixed-wettability surface. Figure 3 gives some snapshots of the boiling processes on the homogeneous hydrophilic surface and the mixed-wettability surface at $t = 4000\delta_t$. The wall superheat is taken as $\Delta T = 0.015$ and the pillar width in the figure varies from $W = 40$ l.u. to 140 l.u. From the figure we can see that the major difference between the boiling processes on the two surfaces lies in that on the homogeneous hydrophilic surface there is no bubble nucleated at the top of the pillar, while on the mixed-wettability surface a bubble is formed at the top of the pillar owing to the hydrophobicity of the top surface. Subsequently, the bubble will interact with the four bubbles nucleated around the square pillar, yielding a larger bubble, which can be observed in the left and middle panels of Fig. 3(b).

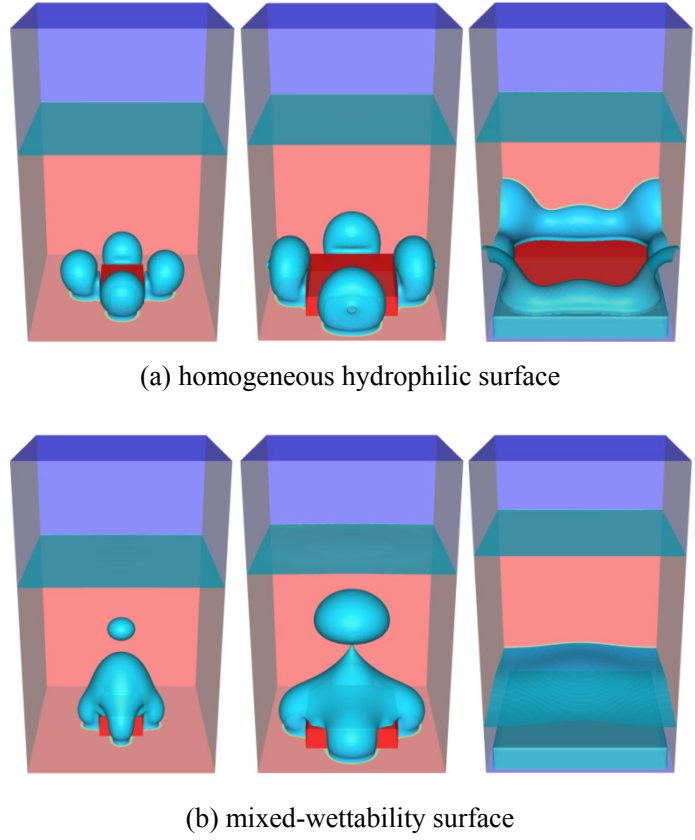

(a) homogeneous hydrophilic surface

(b) mixed-wettability surface

**Fig. 3.** Snapshots of boiling on the homogeneous hydrophilic surface and the mixed-wettability surface at $\Delta T = 0.015$ and $t = 4000\delta_t$. From left to right, the pillar width is $W$ = 40 l.u., 80 l.u., and 140 l.u., respectively.

For the boiling processes on the mixed-wettability surface, Fig. 4 depicts the variations of the normalized heat flux with the pillar width in the cases of $\Delta T = 0.014$ and 0.015. In the present work, the heat flux is the time and spatial average of the local transient heat flux during $2 \times 10^4$ time steps, and then the normalized heat flux is obtained via $\bar{h} = h\Delta x/(\lambda_V T_c)$, where $\Delta x$ is the grid spacing, $T_c$ is the critical temperature, and $\lambda_V$ is the thermal conductivity of the vapor phase. Such a definition of the normalized heat flux is similar to that used in the study of Gong and Cheng [54]. Figure 4 shows that in the case of $\Delta T = 0.015$ the heat flux initially increases with the increase of the pillar width and reaches its peak at $W$ = 80 l.u. After that, the heat flux gradually decreases. A similar trend can also be observed in the case of $\Delta T = 0.014$. Such a trend is attributed to the fact that increasing the pillar width can promote the bubble nucleation at the four edges formed by the bottom substrate and the square pillar.

However, when the pillar width is further increased, the boiling process may enter into the transition or film boiling, which results in a sharp decrease of the heat flux, as shown in Fig. 4 for $W = 140$ l.u.

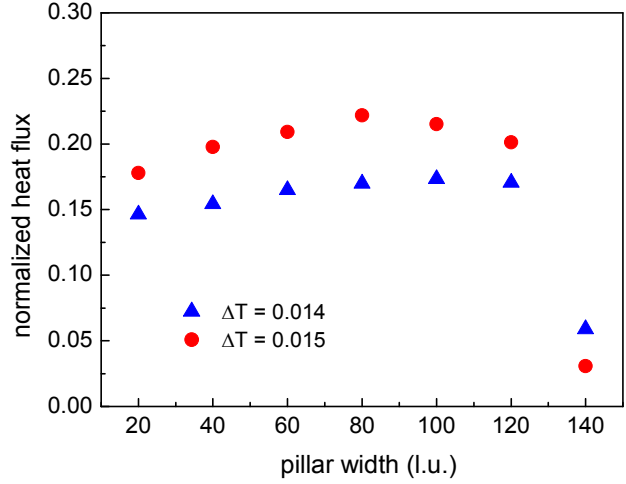

**Fig. 4.** Simulations of boiling on the mixed-wettability surface. Variations of the normalized heat flux with the pillar width in the cases of $\Delta T = 0.014$ and 0.015, respectively.

Now we turn our attention to the boiling performance on the improved mixed-wettability surface, in which the top of the pillar is partially subjected to wettability modification. According to Fig. 4, the best boiling performance on the mixed-wettability surface is achieved around $W = 80$ l.u. in the cases of $\Delta T = 0.014$ and 0.015. For comparison, the pillar width of the improved mixed-wettability surface is also chosen as $W = 80$ l.u. A square wettability-modified region is located at the center of the pillar top (see Fig. 2(c)). The width of the wettability-modified region is denoted by $W_{mod}$, which is smaller than the pillar width. Figure 5 displays the variations of the normalized heat flux with the width of the wettability-modified region in the cases of $\Delta T = 0.014$, 0.015, and 0.016. From the figure we can observe the same trend for the three cases, namely the heat flux initially increases with the increase of the width of the wettability-modified region, and then gradually decreases after reaching its peak value. More specifically, the peak of the heat flux is achieved at $W_{mod} = 60$ l.u., 50 l.u., and 40 l.u. for the cases of $\Delta T = 0.014$, 0.015, and 0.016, respectively, which implies that a smaller $W_{mod}$ is required when the wall superheat increases. Moreover, Fig. 5 also shows that in each case the peak value of the heat flux is much higher than the heat flux achieved at $W_{mod} = 80$ l.u., indicating that the improved mixed-wettability surface performs better than the mixed-wettability surface (i.e., $W_{mod} = W = 80$ l.u.). For example, in the case of $\Delta T = 0.015$ the heat flux is increased by 31% when $W_{mod}$ varies from 80 l.u. to 50 l.u.

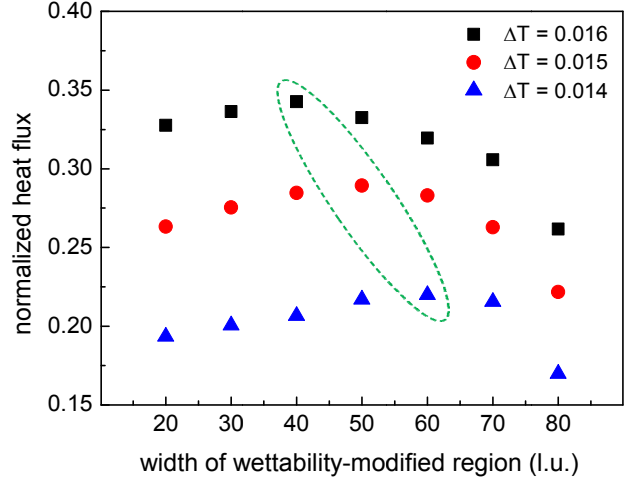

**Fig. 5.** Simulations of boiling on the improved mixed-wettability surface, in which the top of the pillar is partially subjected to wettability modification (denoted by the light blue region in Fig. 2(c)). Variations of the normalized heat flux with the width of the wettability-modified region (i.e., $W_{mod}$) in the cases of $\Delta T = 0.014$, 0.015, and 0.016.

In order to reveal the joint effects of the square pillar (surface structure) and the mixed wettability, the bubble dynamics is analyzed by taking the case of $\Delta T = 0.015$ as an example. Figure 6 displays some snapshots of the boiling process on the improved mixed-wettability surface with $W_{mod} = 50$ l.u. To illustrate the bubble dynamics more clearly, we have plotted two more square pillars in Fig. 6. Firstly, we can observe that the bubbles are nucleated at two regions, namely the four edges formed by the square pillar and the bottom substrate, and the wettability-modified region on the top of the pillar. The former arises from the surface structure, while the latter is generated by the wettability modification. As time goes by, the bubbles gradually grow up. Particularly, for the bubbles nucleated around the square pillar, each bubble will coalesce with another bubble generated from an adjacent pillar (see Fig. 6(b)), and then grows up and departs from the heating surface, as shown in Fig. 6(c). In contrast, the bubble nucleated on the top of the pillar does not coalesce with the bubbles on the bottom substrate.

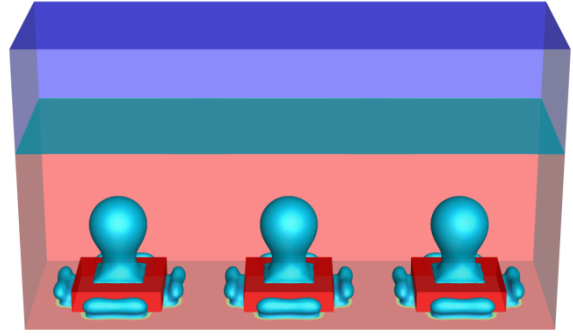

(a) $t = 3000\delta_t$

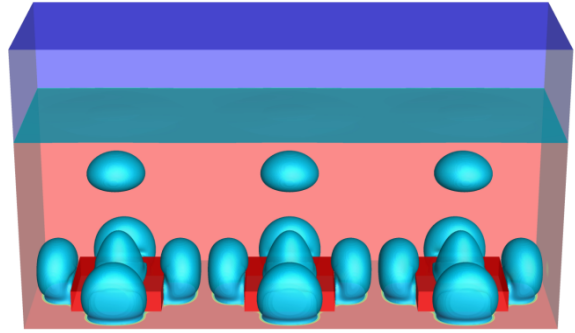

(b) $t = 4500\delta_t$

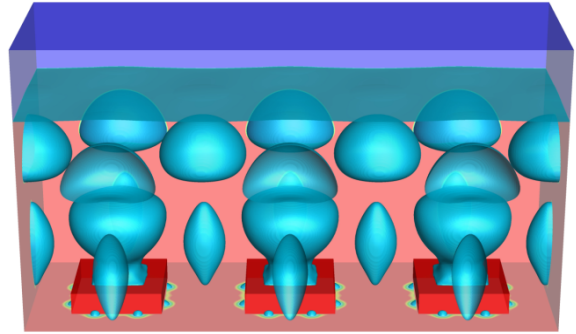

(c) $t = 9500\delta_t$

**Fig. 6.** Snapshots of boiling on the improved mixed-wettability surface in the case of $\Delta T = 0.015$. The pillar width is $W = 80$ l.u., while the width of wettability-modified region is $W_{mod} = 50$ l.u. Two more pillars are plotted in this figure so as to illustrate the bubble dynamics more clearly.

Actually, for the bubble on the pillar top, the three-phase contact lines are pinned at the hydrophilic-hydrophobic boundary, which can be seen in Fig. 6. The size of the bubble on the pillar top is limited by the width of the wettability-modified region and the vapor expansion is constrained by the hydrophilic-hydrophobic boundary. As shown in Fig. 6 ($W_{mod} = 50$ l.u.), the bubble on the pillar top does not interfere with the coalescence-departure mechanism of the bubbles nucleated around the square pillar.

In other words, the expansion of the bubble on the pillar top and its coalescence with the bubbles around the square pillar have been interrupted by the hydrophilic-hydrophobic boundary. However, when the width of the wettability-modified region is further increased, the size of the bubble on the pillar top would increase. Correspondingly, the bubble on the pillar top may coalesce with the bubbles nucleated around the pillar, and then the coalescence-departure mechanism of the bubbles on the bottom substrate may be affected.

To illustrate the aforementioned point, some snapshots of the boiling process on the mixed-wettability surface ($W_{mod} = W = 80$ l.u.) are shown in Fig. 7. By comparing the bubble dynamics in Fig. 7 with that in Fig. 6, we can find that in Fig. 7 the bubbles nucleated around the square pillar no longer coalesce with the bubbles generated from adjacent pillars. As a result, the multiple bubble-departure mode in Fig. 6 is changed to the single bubble-departure mode in Fig. 7, which definitely affects the boiling heat transfer. Figure 8 illustrates the variations of the normalized transient heat flux with time during the boiling processes on the mixed-wettability surface and the improved mixed-wettability surface ($W_{mod} = 50$ l.u.) in the case of $\Delta T = 0.015$. As shown in the figure, the transient heat flux obtained by the improved mixed-wettability surface is much higher than that of the mixed-wettability surface.

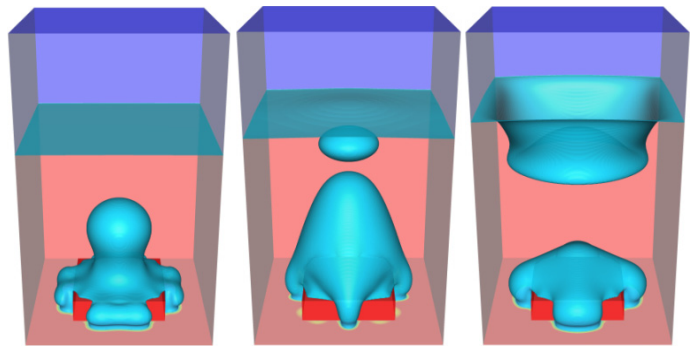

**Fig. 7.** Snapshots of boiling process on the improved mixed-wettability surface with $W_{mod} = W = 80$ l.u. in the case of $\Delta T = 0.015$, which reduces to the mixed-wettability surface. From left to right: $t = 3000\delta_t$, $5000\delta_t$, and $8000\delta_t$, respectively.

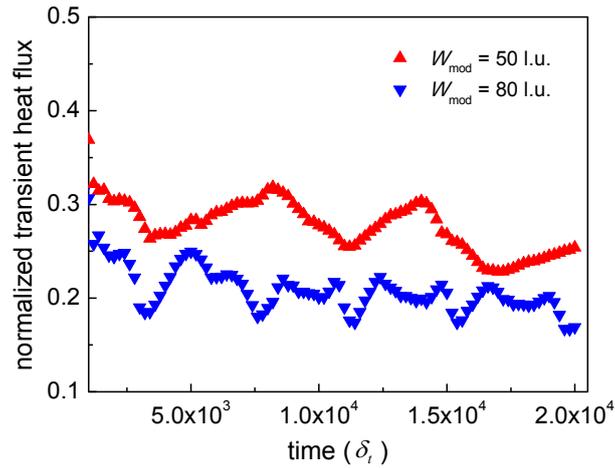

**Fig. 8.** Variations of the normalized transient heat flux with time during the boiling processes on the improved mixed-wettability surface ($W_{mod}$ = 50 l.u.) and the mixed-wettability surface ($W_{mod}$ = $W$ = 80 l.u.) in the case of $\Delta T = 0.015$.

### 3.3. Influences of the shape of the wettability-modified region

In this subsection, we numerically investigate the influences of the shape of wettability-modified region on the boiling performance of the improved mixed-wettability surface. Three different shapes are studied: square, circle, and 45°-rotating square, which are illustrated in Fig. 9(a), 9(b), and 9(c), respectively. The definition of the width of the wettability-modified region is also illustrated in Fig. 9. For the three shapes, the area of the wettability-modified region is given by, respectively

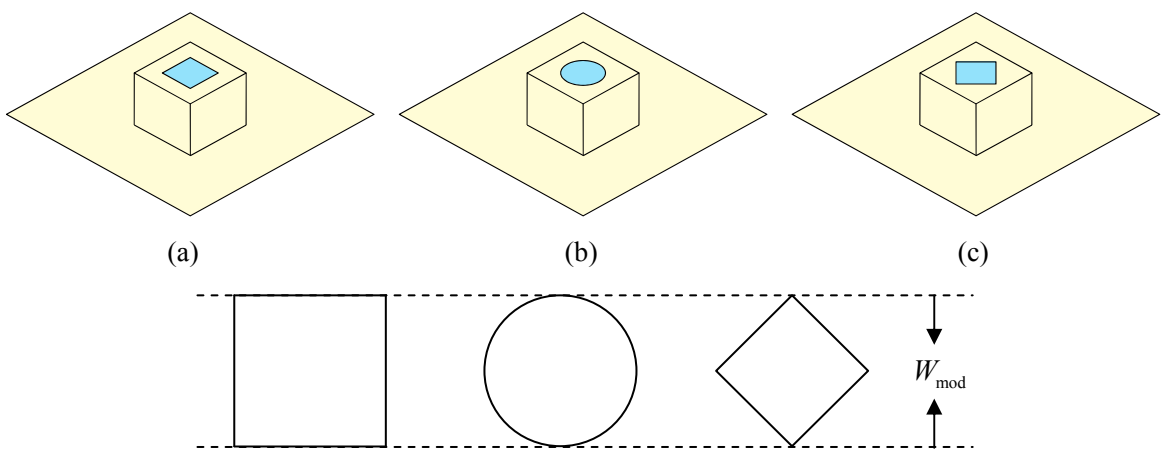

**Fig. 9.** Three different shapes of the wettability-modified region. (a) square, (b) circle, and (c) 45°-rotating square.

$$S_{\text{square}} = W_{\text{mod}}^2, \quad S_{\text{circle}} = \frac{\pi}{4} W_{\text{mod}}^2, \quad S_{\text{square, 45°}} = \frac{1}{2} W_{\text{mod}}^2. \tag{12}$$

For a given value of $W_{\text{mod}}$, it can be found that $S_{\text{square}} > S_{\text{circle}} > S_{\text{square, 45°}}$.

The simulation parameters are the same as those used in the previous subsection and the contact angle of the wettability-modified region is still chosen as $\theta_{\text{pho}} \approx 94°$. Figure 10(a) shows the variations of the normalized heat flux with the width of the wettability-modified region (i.e., $W_{\text{mod}}$) when different shapes are applied. The wall superheat is $\Delta T = 0.015$. From Fig. 10(a) the following phenomena can be observed. Firstly, we can see that the heat fluxes in the cases of circle and 45°-rotating square show the same trend as that in the case of square, i.e., the heat flux initially increases with the increase of $W_{\text{mod}}$, and then gradually declines after reaching its peak value. The reason for this trend has been previously discussed. Furthermore, it can be found that the peak of the heat flux is achieved at $W_{\text{mod}}$ = 50 l.u., 50 l.u., and 60 l.u. in the cases of square, circle, and 45°-rotating square, respectively. When $W_{\text{mod}} \leq 50$ l.u., the heat flux in the case of square is higher than that in the case of circle, which is in turn higher than the heat flux in the case of 45°-rotating square. Such a phenomenon is attributed to the fact that the area of wettability-modified region satisfies the relationship $S_{\text{square}} > S_{\text{circle}} > S_{\text{square, 45°}}$ for a given $W_{\text{mod}}$.

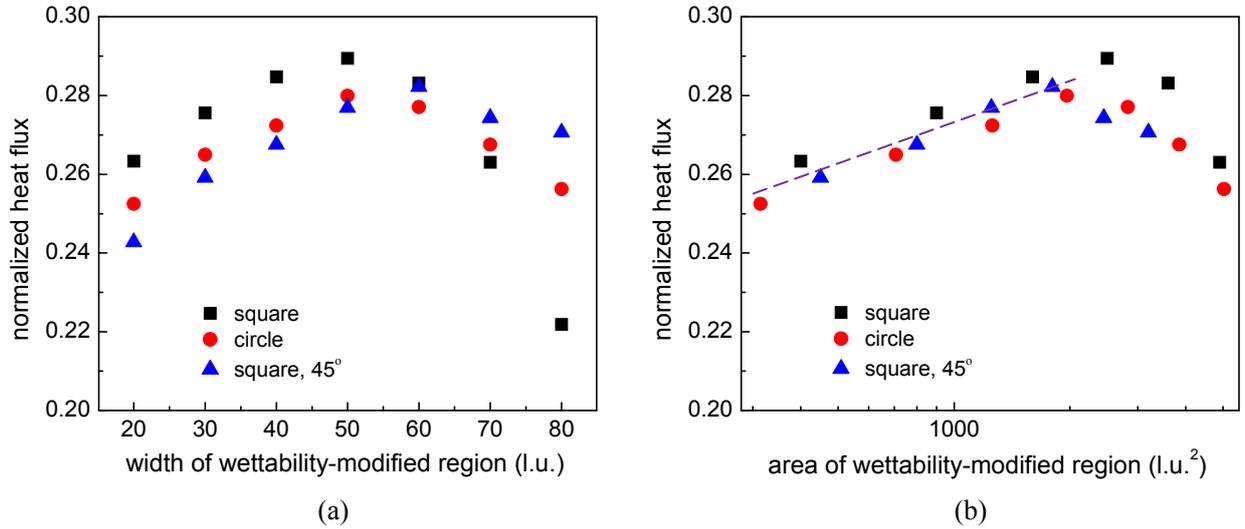

Fig. 10. Simulations of boiling on the improved mixed-wettability surface with different shapes (square, circle, and 45°-rotating square) of the wettability-modified region. Variations of the normalized heat flux with (a) the width of the wettability-modified region and (b) the area of the wettability-modified region, respectively. The wall superheat is $\Delta T = 0.015$.

The variations of the normalized heat flux with the area of the wettability-modified region are depicted in Fig. 10(b). In light of the area of the wettability-modified region, we can find that, before reaching the peak values, the heat fluxes of the three cases basically increase along the dotted line plotted in Fig. 10(b) and there are no significant differences between the results of the three cases. This is because in these situations the bubble nucleated on the pillar top does not coalesce with the bubbles on the bottom substrate and hence the heat flux increases with increasing the area of the wettability-modified region. Contrarily, after reaching the peak value, the heat flux would decrease with the increase of the area of the wettability-modified region because the bubble nucleated on the pillar top will affect the coalescence-departure mechanism of the bubbles on the bottom substrate.

To illustrate the phenomenon that the heat flux peaks at $W_{\text{mod}} = 50$ l.u. in the cases of square and circle, but peaks at $W_{\text{mod}} = 60$ l.u. in the case of 45°-rotating square, in Figs. 11 and 12 we display some snapshots of the boiling processes on the improved mixed-wettability surface when $W_{\text{mod}}$ is taken as 50 l.u. and 60 l.u., respectively. The left, middle, and right panels of the figures show the results at $t = 2000\delta_t$, $4000\delta_t$, and $6000\delta_t$, respectively. The left panels of these two figures clearly show that the size of the bubble nucleated on the pillar top increases with increasing the area of the wettability-modified region. Since the area satisfies the relationship $S_{\text{square, 45°}} < S_{\text{circle}} < S_{\text{square}}$ for a given $W_{\text{mod}}$, the bubble on the pillar top is smallest in the case of 45°-rotating square.

Particularly, from Figs. 11(c) and 12(c) we can see that in the case of 45°-rotating square the bubble on the pillar top does not interfere with the coalescence-departure mechanism of the bubbles on the bottom substrate during the entire boiling process. However, by comparing the right panel of Fig. 12 with that of Fig. 11, we can find that in the right panels of Figs. 12(a) and 12(b) (the cases of square and circle) the bubble on the pillar top has coalesced with the bubbles on the bottom substrate and a large portion of the pillar top has been covered by vapor. Such a phenomenon can also be observed in Fig. 13, which displays the density contours of the middle $x$-$z$ cross section of the results shown in the right panel of Fig. 12. Accordingly, when the width of the wettability-modified region in Fig. 10(a) is increased from 50 l.u.

to 60 l.u., the heat fluxes in the cases of square and circle decrease, whereas the heat flux in the case of 45°-rotating square increases. Nevertheless, since the area of the wettability-modified region in the case of square with $W_{mod}$ = 50 l.u. is larger than the area in the case of 45°-rotating square with $W_{mod}$ = 60 l.u., a relatively higher heat flux is yielded in the case of square with $W_{mod}$ = 50 l.u., as shown in Fig. 10(a).

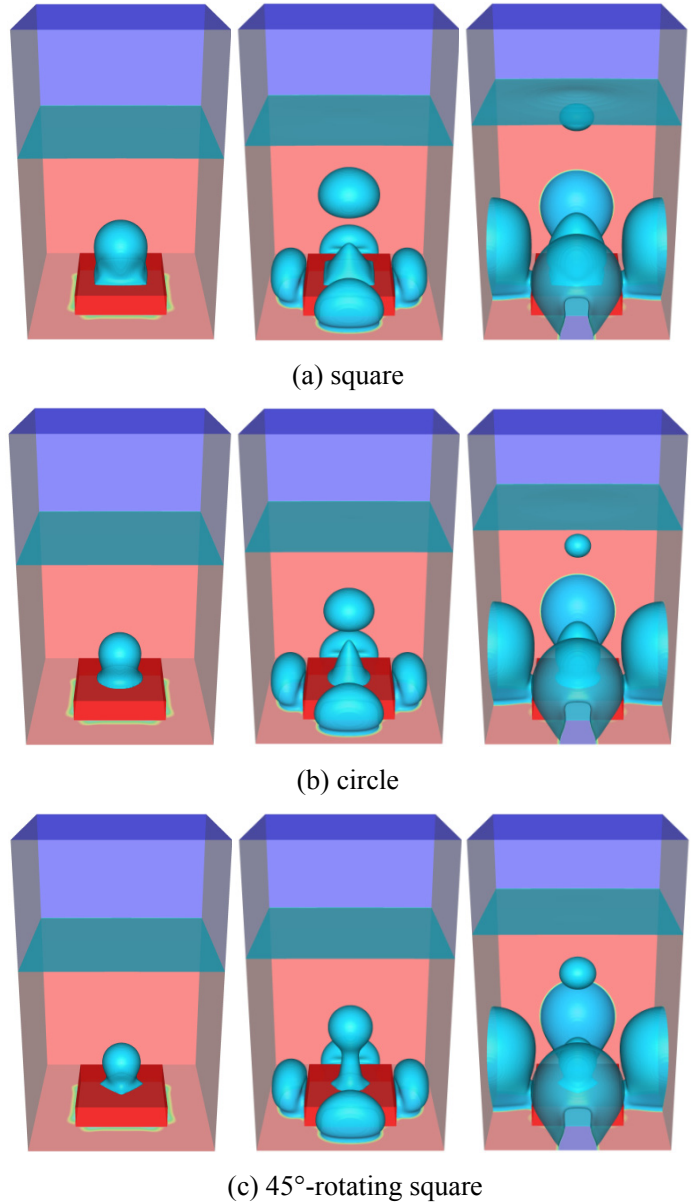

(a) square

(b) circle

(c) 45°-rotating square

**Fig. 11.** Snapshots of boiling on the improved mixed-wettability surface with different shapes of the wettability-modified region. The width of wettability-modified region is $W_{mod}$ = 50 l.u. and the wall superheat is $\Delta T = 0.015$. From left to right: $t = 2000\delta_t$, $4000\delta_t$, and $6000\delta_t$, respectively.

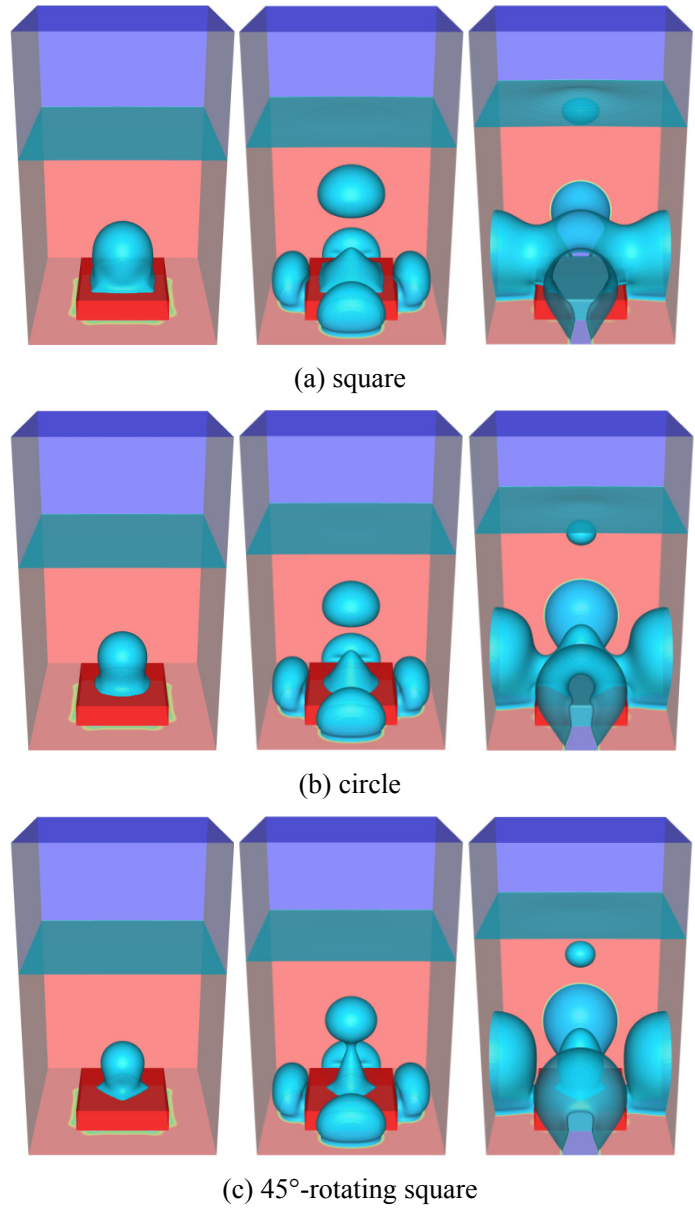

**Fig. 12.** Snapshots of boiling on the improved mixed-wettability surface with different shapes of the wettability-modified region. The width of wettability-modified region is $W_{\text{mod}} = 60$ l.u. and the wall superheat is $\Delta T = 0.015$. From left to right: $t = 2000\delta_t$, $4000\delta_t$, and $6000\delta_t$, respectively.

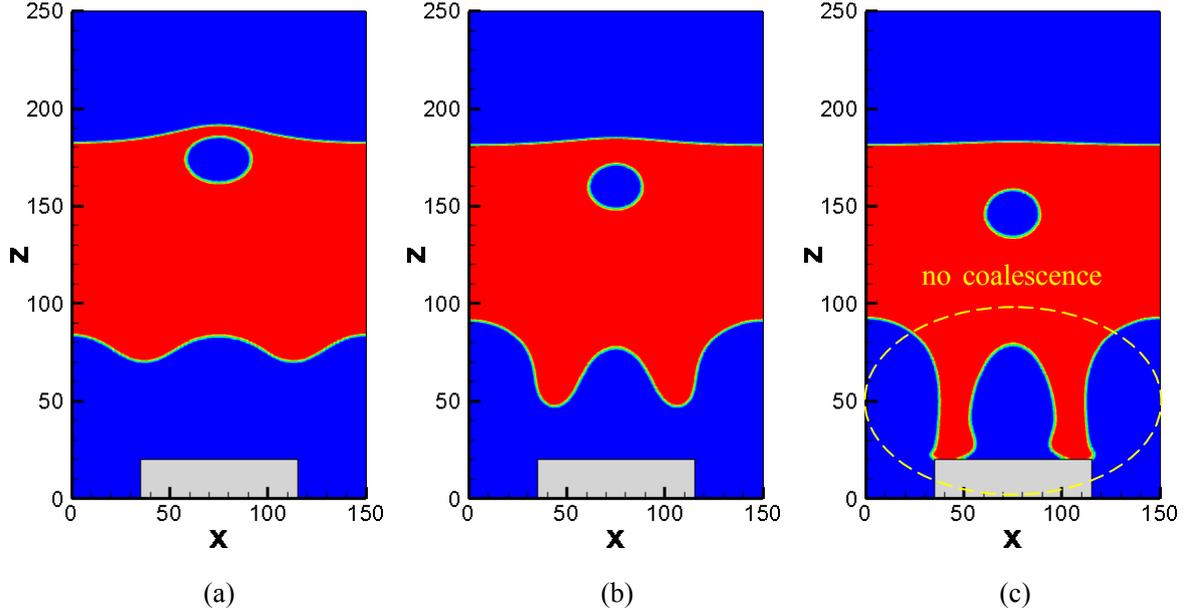

(a)                (b)                (c)

**Fig. 13.** Density contours of the *x-z* cross section at $y = L_y/2$ of the results shown in the right-hand panel of Fig. 12. From left to right: the cases of (a) square, (b) circle, and (c) 45°-rotating square, respectively. The blue and red regions represent the vapor and liquid phases, respectively, while the light gray region is the square pillar.

To further illustrate the aforementioned points, Fig. 14 compares the normalized transient heat fluxes in the cases of $W_{mod}$ = 50 l.u. and 60 l.u. when a square wettability-modified region is applied. In the figure the two snapshots were taken for the case of $W_{mod}$ = 60 l.u. at $t = 5000\delta_t$ and $7000\delta_t$, respectively. From the figure we can see that in the early stage the transient heat flux of the case $W_{mod}$ = 60 l.u. is higher than that of the case $W_{mod}$ = 50 l.u., owing to the fact that the area of the wettability-modified region in the former case is larger than that in the latter case. However, apparent differences can be observed between the transient heat fluxes of these two cases after $t \approx 4500\delta_t$, from which the bubble on the pillar top in the case of $W_{mod}$ = 60 l.u. starts to interfere with the coalescence-departure mechanism of the bubbles on the bottom substrate. To sum up, the numerical results presented in this subsection demonstrate that the shape of the wettability-modified region has an important influence on the boiling performance of the improved mixed-wettability surface and also further confirm the analysis in the previous subsection regarding the joint effects of surface structure and mixed wettability for enhancing nucleate boiling heat transfer. Generally, it is observed that the square performs best among the three investigated shapes since it provides the highest value of the heat flux (see Fig. 10), but its heat flux is relatively sensitive to the width of the wettability-modified region in comparison with the other two shapes.

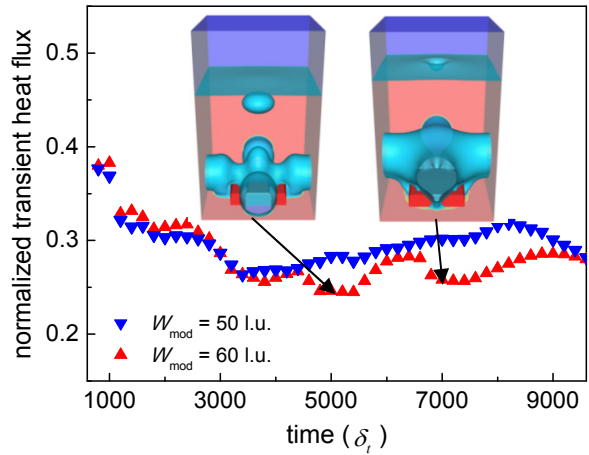

**Fig. 14.** Variations of the normalized transient heat flux with time during the boiling processes on the improved mixed-wettability surface with the wettability-modified region being a square. A comparison is made between the cases of $W_{\mathrm{mod}}$ = 50 l.u. and 60 l.u.

## 4. Conclusions

The combination of microstructures and mixed wettability has been applied to enhance nucleate boiling heat transfer in some recent experimental and numerical studies. However, in the existing studies, the tops of microstructures were entirely subjected to wettability modification, which makes the effect of mixed wettability dependant on the characteristic length of microstructures. In order to disclose the joint effects of surface structure and mixed wettability on nucleate boiling, in this paper we have designed an improved type of pillar-textured surface with mixed wettability, in which the tops of square pillars are *partially* subjected to wettability modification. Numerical investigation of the boiling heat transfer performance on the improved mixed-wettability surface has been performed using a 3D thermal multiphase LB model. The main findings and conclusions are summarized as follows.

(i) For the improved mixed-wettability surface, bubbles are nucleated at two regions: one is the four edges formed by the square pillar and the bottom substrate, which is caused by the surface structure, and the other is the wettability-modified region on the top of the pillar, which is yielded by the wettability modification.

(ii) It is found that the width of the wettability-modified region plays an important role in the boiling performance of the improved mixed-wettability surface and the best boiling performance is achieved in the situation that the width of the wettability-modified region is sufficiently large but the bubble nucleated on the pillar top still does not interfere with the coalescence-departure mechanism of the bubbles nucleated around the pillar, which takes the advantage of the multiple bubble-departure mode and optimizes the joint effects of surface structure and mixed wettability for enhancing nucleate boiling heat transfer.

(iii) The influences of the shape of the wettability-modified region have also been studied. Three different shapes have been considered: square, circle, and 45°-rotating square. Numerical results show that the square performs best among the investigated shapes as it gives the highest value of the heat flux, but its heat flux is relatively sensitive to the width of the wettability-modified region in comparison with the other two shapes.

## Acknowledgments

This work was supported by the National Natural Science Foundation of China (Grant No. 51822606).

## Appendix

In this appendix, the code verification and model validation are made through verifying the well-known $D^2$ law [55, 56]. Such a law is established for droplet vaporization based on the following conditions: the liquid and vapor phases are quasi-steady, the thermo-physical properties are constant, and the liquid-vapor phase change occurs in an environment with negligible viscous heat dissipation and no buoyancy [29]. Numerical simulations are carried out in a square domain with $L_x \times L_y \times L_z = 150 \text{ l.u.} \times 150 \text{ l.u.} \times 150 \text{ l.u.}$ Initially, a droplet with a diameter of $D_0 = 50 \text{ l.u.}$ is located at the center of the computational domain. At the initial time step, the temperature of the droplet is the saturation temperature $T_s = 0.86 T_c$ and the temperature of the surrounding vapor is $T_s + \Delta T$, in which the superheat is chosen as $\Delta T \approx 0.015$. The kinematic viscosity is taken as $\nu = 0.1$ and the specific heat at constant volume is chosen as $c_V = 5$. Two cases are considered for the thermal conductivity, i.e., case A with $\lambda = 1/3$ and case B with $\lambda = 2/3$. As the simulation starts, the droplet will gradually evaporate owing to the temperature gradient at the liquid-vapor interface. According to the $D^2$ law, the evaporation rate depends linearly on $\lambda$. Figure 15 displays the variations of the squared diameter ratio $(D/D_0)^2$ with time for the two cases. As expected, in both cases the squared diameter ratio decreases linearly with time, satisfying the $D^2$ law. Quantitatively, the parameter $k$ in the law of $D^2(t)/D_0^2 = 1 - kt$ is given by $k \approx 0.81 \times 10^{-5}$ and $1.60 \times 10^{-5}$ for cases A and B, respectively. It can be found that the parameter $k$ of case B is about twice as large as that of case A.

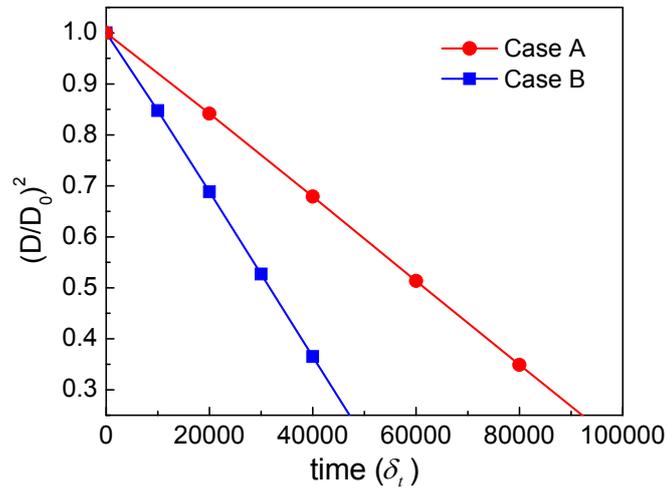

**Fig. 15.** Variations of the squared diameter ratio with time (the $D^2$ law problem).